\documentclass[12pt,a4paper,twoside]{article}
\RequirePackage{latexsym,amsmath,amssymb}
\RequirePackage[dvipsnames,usenames]{color}
\usepackage{a4wide}
\usepackage{amssymb}
\usepackage{bm}
\usepackage{dcolumn}
\usepackage{amsfonts}
\usepackage{graphicx,epsfig}
\usepackage{psfrag}
\usepackage{multicol}
\usepackage{tikz}
\usetikzlibrary{arrows,shapes}
\usetikzlibrary{matrix,arrows}
\newcommand{\be}{\begin{eqnarray}}
\newcommand{\ee}{\end{eqnarray}}

\newcommand{\ket}[1]{\mbox{$\mid #1\,\rangle$}}

\newcommand{\pro}[2]{\mbox{$\langle\, #1 \mid #2\,\rangle$}}
\newcommand{\expec}[1]{\mbox{$\langle\, #1\,\rangle$}}

\renewcommand{\d}{\mbox{${\rm d}$}} 
\newcommand{\lp}{\ell_{\rm p}}
\newcommand{\mpl}{m_{\rm p}}

\newcommand{\rh}{r_{\rm H}}
\newcommand{\Rh}{R_{\rm H}}
\newcommand{\psis}{{\psi}_{\rm S}}

\newcommand{\psih}{{\psi}_{\rm H}}
\begin{document}
 
\begin{titlepage}
%
%
%
\title{\bf The horizon of the lightest black hole}
\author{Xavier~Calmet\thanks{E-mail: x.calmet@sussex.ac.uk} $^a$ ~and~
Roberto~Casadio\thanks{E-mail: casadio@bo.infn.it} $^{b,c}$
\\
\\
{\em $^a$ Physics $\&$ Astronomy, University of Sussex}   
\\
{\em Falmer, Brighton, BN1 9QH, United Kingdom }
\\
\\
{\em $^b$Dipartimento di Fisica e Astronomia, Universit\`a di Bologna}
\\
{\em via Irnerio~46, I-40126 Bologna, Italy}
\\
\\
{\em $^c$I.N.F.N., Sezione di Bologna,}
\\
{\em via B.~Pichat~6/2, I-40127 Bologna, Italy}
}
%
%
\maketitle
\begin{abstract}
We study the properties of the poles of the resummed graviton propagator obtained by
resumming bubble matter diagrams which correct the classical graviton propagator.
These poles have been previously interpreted as black holes precursors.
Here, we show using the Horizon Wave-Function formalism that these poles indeed
have properties which make them compatible with being black hole precursors.
In particular, when modeled with a Breit-Wigner distribution, they have a well defined
gravitational radius.
The probability that the resonance is inside its own gravitational radius, and thus that
it is  black hole, is about one half.
Our results confirm the interpretation of these poles as black hole precursors.
\end{abstract}
\end{titlepage}
\newpage
\section{Introduction}
\label{intro}
\setcounter{equation}{0}
The aim of this paper is to investigate further the properties of the black holes precursors
that have been identified in~\cite{calmet14} using an effective theory approach for gravity
and resummation techniques.
In particular, we shall study whether these objects have an horizon and can thus truly
be identified with black holes. 
\par
Obviously, quantum black holes are quantum gravitational objects, but while we are still
far from having a theory of quantum gravity, effective field theory techniques can be reliably
applied to General Relativity coupled to matter at energy scales below the energy scale at
which quantum gravitational effects become of the
same magnitude as quantum effects generated by the other forces of
nature~\cite{Donoghue:2012zc,Donoghue:1994dn,Calmet:2013hfa,Calmet:2014sfa}. 
The leading order terms of the effective field theory are given by
\begin{eqnarray}
\label{action1}
S
=
\int d^4x \, \sqrt{-g} \left[
\frac{\mpl^2}{16 \pi}\, \mathcal{R}(g)
+ c_1\, \mathcal{R}(g)^2
+ c_2\, \mathcal{R}_{\mu\nu}(g)\,\mathcal{R}^{\mu\nu}(g) 
+ \mathcal{L}_{SM}
+ \mathcal{O}(\Lambda_c) 
\right]
\ , 
\end{eqnarray}
where $ \mathcal{R}(g)$ is the Ricci scalar, $\mathcal{R}^{\mu\nu}(g)$ is the Ricci
tensor, the metric $g_{\mu\nu}$ describes the graviton when the action is
linearized, and $\mathcal{L}_{SM}$  stands for the Lagrangian of the standard model
of particle physics.
The action contains two energy scales, the Planck scale $\mpl=1.2209 \times10^{19}\,$GeV which is related to Newton's constant by $\mpl=1/\sqrt{G_N}$
and a scale $\Lambda_c$ which is the energy scale at which we expect the effective
field theory to break down.
The constants $c_1$ and $c_2$ are dimensionless ones.
We have suppressed the cosmological constant and a potential non-minimal coupling
of the Higgs boson to the Ricci scalar which are not important for our considerations.
It is important to realize that the two scales $\mpl$ and $\Lambda_c$ need not to be identical.
The Planck scale  is the gravitational coupling constant
which appears in the vertices of Feynman diagrams which involve gravitons.
The other dimensionful parameter of the model, the cut-off of the effective field theory,
$\Lambda_c$ is related to the Planck scale, but as we shall see shortly, it has recently
been shown to be dependent on the number of fields in the matter sector~\cite{calmet14}.
\par
Working in linearized General Relativity and in a Minkowski background, it is possible to
resum loop diagrams involving matter fields which correct the graviton's propagator.
This correction is calculated~\cite{Aydemir:2012nz} in the large $N$ limit, where
$N=N_s +3 N_f +12 N_V$ ($N_s$, $N_f$ and $N_V$ are respectively the number of
real scalar fields, fermions and spin~1 fields in the model), while keeping $N\,G_N$ small.
One uses dimensional analysis to regulate the integrals and absorb the divergent parts
of the diagrams into the coefficients of $R^2$ and $R_{\mu\nu}\,R^{\mu\nu}$.
Note that in the standard model $N_s=4$, $N_f=45$ and $N_V=12$, so $N=283$.
In other words there are many more matter degrees of freedom than gravitational ones
(we assume that there is only one massless graviton).
Loops involving the graviton are thus suppressed by factors of $1/N$ compared to matter
loops (at least as long as one considers energies below the Planck scale) and perturbation
theory can be trusted.
\par
This large $N$ resummation leads to resummed graviton propagator given by \cite{Aydemir:2012nz}
\begin{eqnarray}
\label{resprop}
i\,D^{\alpha \beta,\mu\nu}(q^2)
=
i \left (L^{\alpha \mu}L^{\beta \nu}+L^{\alpha \nu}L^{\beta \mu}-L^{\alpha \beta}L^{\mu \nu}\right)
\Delta(q^2)
\ ,
\end{eqnarray}
with $L^{\mu\nu}(q)=\eta^{\mu\nu}-q^\mu q^\nu /q^2$ and
\be
\Delta(q^2)
=
\frac{1}
{2\,q^2\left [1 - \frac{N\, q^2}{120\,\pi\,\mpl^2} \log \left (-\frac{q^2}{\mu^2} \right) \right]}
\,
\ee
where $\mu$ is the renormalization scale.
This resummation was first considered in when studying the perturbative unitarity of the
effective action~\eqref{action1}~\cite{Aydemir:2012nz,Han:2004wt,Atkins:2010eq,Atkins:2010re,Atkins:2010yg,Antoniadis:2011bi,Calmet:2013hia}.
\par
In~\cite{calmet14}, it has been proposed to to interpret the massive poles of this propagator
as Planck-size black hole precursors or quantum black holes.
The position of the poles determines the mass and the width of the precursors:
$p_0^2=(M_{\rm BH}+ i\,\Gamma_{\rm BH}/2)^2$. 
The poles of the resummed propagator~\eqref{resprop} are given by
\begin{eqnarray}
q^2_{1}
&=&
0
\nonumber
\\
\label{poles}
\\
q^2_{2}
&=&
(q^2_{3})^*
=
\frac{\frac{120\,\pi}{N}\,\mpl^2}{ W\!\left (-\frac{120\,\pi}{N}\,\frac{\mpl^2}{\mu^2}\right)}
=\left(M_{\rm BH}+ i\, {\Gamma_{\rm BH}}/{2}\right)^2
\ ,
\nonumber
\end{eqnarray}
where $W(x)$ is the Lambert W-function.
The pole at $q^2=0$ corresponds to the usual massless graviton.
The position of the pole and hence the energy scale at which non-perturbative effects
are becoming important depends on the matter content of the model, i.e.~on $N$.
As mentioned above, in the standard model one has $N=283$ and the complex pole at
$q^2=q^2_{2}$ corresponds to a particle with mass~\cite{calmet14}
\be 
M_{\rm BH}
\simeq 
7.2\times 10^{18}\,{\rm GeV}
\simeq
\sqrt{\frac{120\,\pi}{N}}\,\frac{\mpl}{2}
\ ,
\ee
and width
\be
\Gamma_{\rm BH}
\simeq
6.0\times 10^{18}\,{\rm  GeV}
\simeq
\sqrt{\frac{120\,\pi}{N}}\,\frac{\mpl}{2}
\ .
\ee
As explained in~\cite{calmet14}, the mass and the width of the lightest
of black holes depends on the parameter $N$.
It is natural to interpret these poles as black hole precursors or non-local
extended objects since the resummed propagator leads to non-local effects
in gravity~\cite{Donoghue:2014yha} and quantum field theory~\cite{Calmet:2015dpa}. 
This interpretation is also compatible with generic
arguments~\cite{Mead:1964zz,Garay:1994en,Padmanabhan:1987au,Calmet:2004mp} 
based on quantum mechanics and general relativity which lead to the notion of a minimal
length and thus some kind of non-locality. 
Obviously, these estimates depend on the renormalization scale which is taken
of the order of the Planck mass. One can use the spectral decomposition to write the
propagator as
\be
\Delta(q^2)
=
\frac{1}{q^2}
+
\frac{R_2}{q^2-q_2^2}
+
\frac{R_3}{q^2-q_3^2}
+
\int_{M_{\rm BH}^2}^\infty
ds\,\frac{\rho(s)}{s-q^2}
\ ,
\label{spectral}
\ee
where $R_{2/3}$ are the residues at the two non-trivial poles.
The second complex pole at $q^2=q_3^2$ would leads to acausal effects.
Several mechanisms could eliminate this pole
(see, e.g.~\cite{Donoghue:2014yha,Espriu:2005qn,Barvinsky:1985an,Barvinsky:1995jv,Barvinsky:1994ic,Avramidi:1990je}
where the log-term is reinterpreted as a non-local interpolating function which leads
to causal effects).
However, we shall assume that this is the scale above which we cannot trust perturbation
theory in the standard model.
\par
The effective field theory does not provide reliable information about the spectral
density function $\rho(s)$.
However, we have some information about this function coming from black hole physics.
We expect the classical regime to begin around 5 to 20 times the mass of the first black
hole (see e.g.~\cite{Calmet:2014dea}).
At that scale, we expect to have a continuum since semi-classical black holes are
expected to have a continuous mass spectrum.
Between $M_{\rm BH}$ and $(5-20) \times M_{\rm BH}$, the situation is more difficult.
In~\cite{Calmet:2014uaa}, it was argued that the mass spectrum of quantum black holes
needs to be quantized, otherwise their virtual effects could lead to large effects in low
energy experiments such as measurements of the anomalous magnetic moment of the muon.
We will assume that $\rho(s)$ is discrete between $M_{\rm BH}$ and the continuous,
semi-classical region.
We assume that the resonances are sharply peaked and do not overlap much.
We shall require that the spacing between the first quantum excitation which we identified
as a pole of the resummed propagator and the next excitation is larger than the width
of the black hole precursor.
In that case, we should be able to trust the model up to a scale
\be
\Lambda_c \simeq
\sqrt{\frac{120\,\pi}{N}}\,\mpl
\simeq
1.4\times 10^{19}\,{\rm GeV}
\ ,
\ee
which corresponds to twice the width of the black hole precursor. 
In other words, we model the mass spectrum between $M_{\rm BH}$ and the continuum
and require that we can trust our model up to the scale $\Lambda_c$ which we take
to be the cut-off for our model of quantum black holes.
\section{Horizon wave-function}
\setcounter{equation}{0}
\label{sHWF}
Our knowledge of black holes in general relativity suggests that these objects are
states somewhat similar to hadrons in QCD, except that gravity democratically confines
all sorts of particles above some critical scale, rather than just strongly interacting ones.
This should be particularly true for quantum black holes~\cite{Calmet:2008dg,Calmet:2011ta}.
It is therefore very likely that, although their existence can be inferred within perturbation
theory, like we have recalled in the previous section, a full description of their quantum
properties requires a non-perturbative approach, like the Horizon Wave-Function (HWF)
formalism (for the details, see Refs.~\cite{timeEvo,fuzzyh}; for a similar picture of the black hole
horizon, see Ref.~\cite{davidson}).
\par
This approach assumes the validity of the Einstein equations in the non-perturbative
regime, and amounts to quantizing the Misner-Sharp mass for spherically symmetric
sources, $m(r,t)
=
4\,\pi\int_0^r \rho(\bar r,t)\,\bar r^2\,\d \bar r$,
which in turn defines the gravitational radius of the system,
\be
\Rh
=
2\,\lp\,\frac{m}{\mpl}
\ .
\label{hoop}
\ee
The latter then identifies the location of a trapping surface if $\Rh(r,t)= r$.
If this relation holds in the vacuum outside the region where the source is located,
$\Rh$ becomes the usual Schwarzschild radius, and the above argument gives
a mathematical foundation to Thorne's hoop conjecture~\cite{hoop},
which roughly states that a black hole forms when the impact parameter $b$
of two colliding small objects is shorter than $\Rh=2\,\lp\,{E}/{\mpl}$, where $E$
is the total energy in the center of mass frame. 
This classical description becomes questionable for sources of the Planck size or
lighter, since quantum effects may not be neglected.
The Heisenberg principle of quantum mechanics introduces an uncertainty in
the spatial localization of a particle of the order of the Compton-de~Broglie
length, $\lambda_m \simeq \lp\,{\mpl}/{m}$.
Since quantum physics is a more refined description of reality, we could argue that
$\Rh$ only makes sense if $\Rh\gtrsim \lambda_m$ or $m\gtrsim\mpl$.
\par
The HWF formalism starts from decomposing the particle's state into energy
eigenstates,
\be
\ket{\psis}
=
\sum_E\,C(E)\,\ket{\psi_E}
\ ,
\ee
where the sum represents the spectral decomposition in Hamiltonian eigenmodes,
\be
\hat H\,\ket{\psi_E}=E\,\ket{\psi_E}
\ ,
\ee
and $H$ should be specified depending on the system at hand.
The gravitational radius~\eqref{hoop} is then quantized by expressing the
energy $E=m$ in terms of the Schwarzschild radius $\rh$ and define the corresponding
wave-function~\footnote{Note we use the lower letter $\rh$ to distinguish
this quantum variable from the classical Schwarzschild radius $\Rh$.}
\be
\psih(\rh)
=
{\mathcal{N}_{\rm H}}\,C(\rh(E))
\ ,
\ee
whose normalization $\mathcal{N}_{\rm H}$ can be fixed by using the norm
defined by the scalar product
\be
\pro{\psih}{\phi_{\rm H}}
=
4\,\pi
\int_0^\infty
\psih^*(\rh)\,\phi_{\rm H}(\rh)\,\rh^2\,\d \rh
\ .
\label{prod}
\ee
Let us remark that this quantum description of the gravitational radius assumes
that, in the static case, the only relevant degrees of freedom associated with the gravitational
structure of space-time (which classically give rise to trapping surfaces)
are those turned on by the degrees of freedom of the matter source.
This implies that we can just consider ``on-shell'' states, for which Eq.~\eqref{hoop}
holds as an operator equation, and neglect gravitational fluctuations,
which could be studied by employing standard background field method techniques.
\par
The normalized wave-function $\psih$ yields the probability that the gravitational radius
has size $r=\rh$, but this radius is ``fuzzy'', like the energy.
Moreover, having defined the $\psih$ associated with a given $\psis$,
we can also define the conditional probability density that the particle lies
inside its own gravitational radius as
\be
\mathcal{P}_<(r<\rh)
=
P_{\rm S}(r<\rh)\,\mathcal{P}_{\rm H}(\rh)
\ ,
\label{PrlessH}
\ee
where
\be
P_{\rm S}(r<\rh)
=
\int_0^{\rh}
\mathcal{P}_{\rm S}(r)\,\d r
=
4\,\pi\,\int_0^{\rh}
|\psis(r)|^2\,r^2\,\d r
\label{PsR}
\ee
is the usual probability that the system lies within the size $r=\rh$, and
\be
\mathcal{P}_{\rm H}(\rh)
=
4\,\pi\,\rh^2\,|\psih(\rh)|^2
\label{Ph}
\ee
is the probability density that the gravitational radius has size $r=\rh$.
One can also view $\mathcal{P}_<(r<\rh)$ as the probability density that the sphere $r=\rh$
is a trapping surface, so that the probability that the system is a black hole
(of any horizon size), will be obtained by integrating~\eqref{PrlessH}
over all possible values of $\rh$, namely
\be
P_{\rm BH}
=
\int_0^\infty
\mathcal{P}_<(r<\rh)\,\d \rh
\ .
\label{Pbh}
\ee
Note that the Planck mass $\mpl$ and length $\lp$ play a crucial role in the above construction,
since they explicitly appear in the definition of the gravitational radius~\eqref{hoop}.
In the following, we shall assume their standard values.
This is consistent with our effective theory approach since we do not consider corrections
to the coefficient of the Ricci scalar in the effective action.
\subsection{Gravitational radius and uncertainty}
We can now derive the HWF for the non-trivial pole corresponding to a well defined
one-particle state~\eqref{poles}.
For simplicity, we model the lightest black hole using a Breit-Wigner distribution
\be
\psi_{\rm S}^*(E)\,\psi_{\rm S}(E)
\equiv
\rho(E)
=
\frac{\mathcal{N}}{(E^2-M_{\rm BH}^2)^2+M_{\rm BH}^2\,\Gamma_{\rm BH}^2}
\ ,
\label{BW}
\ee
where $\mathcal{N}$ is a normalization factor, and $E<E_c$, with
$E_c$ a cut-off corresponding to the beginning of the continuum
spectrum in Eq.~\eqref{spectral}.
In the following we shall assume for simplicity, and in agreement with~\eqref{spectral},
that there is no other discrete resonance in the spectrum below the cutoff for our model,
so that 
\be
E_c\simeq 2\,M_{\rm BH}\simeq\Lambda_c
\ .
\ee
\par
The corresponding HWF is then obtained by assuming the unnormalized 
HWF $|\tilde\psi_{\rm H}|^2\simeq\rho$, and the corresponding probability density~\eqref{Ph}
then reads
\be
\mathcal{P}_{\rm H}\,\d\rh
&=&
\frac{\mpl^3\,M_{\rm BH}}{4\,\lp^3\,F_0(\gamma_{\rm BH},\Lambda)}
\left[
\frac{\rh^2\,\d\rh}
{\left(\frac{\mpl^2\,\rh^2}{4\,\lp^2}-M_{\rm BH}^2\right)^2
+M_{\rm BH}^2\,\Gamma_{\rm BH}^2}
\right]
\nonumber
\\
&=&
\frac{(x+1)^{\frac{1}{2}}}{x^2+\gamma_{\rm BH}^2}\,\frac{\d x}{F_0(\gamma_{\rm BH},\Lambda)}
\ .
\label{PhR}
\ee
where $F_0(\gamma_{\rm BH},\Lambda)\simeq F_0(0.83,3)$ is a number of order one
(see Appendix~\ref{AppI}).
We can now compute expectations values of powers of the gravitational
radius,
\be
\expec{\hat r_{\rm H}^n}
\simeq
\left(2\,\lp\,\frac{M_{\rm BH}}{\mpl}\right)^n
\frac{F_n(0.83,3)}{F_0(0.83,3)}
\ ,
\ee
from which, in particular, one finds
\be
\expec{\hat r_{\rm H}}
\simeq
1.4\,\lp
\ ,
\ee
and
\be
\expec{\hat r_{\rm H}^2}
\simeq
2.2\,\lp^2
\ ,
\ee
so that the relative uncertainty in the gravitational radius is given by
\be
\sqrt{\left|
\frac{\expec{\hat r_{\rm H}^2}-\expec{\hat r_{\rm H}}^2}
{\expec{\hat r_{\rm H}}^2}
\right|}
\simeq
0.3
\ ,
\ee
which means that the gravitational radius is well-defined for such a quantum object.
\subsection{Black hole probability}
We cannot yet claim the resonance is a black hole.
For that, we need to show that the quantum state of this resonance is located
mostly inside the gravitational radius.
\par
\begin{figure}[t]
\centering
\includegraphics[width=7.5cm,height=5cm]{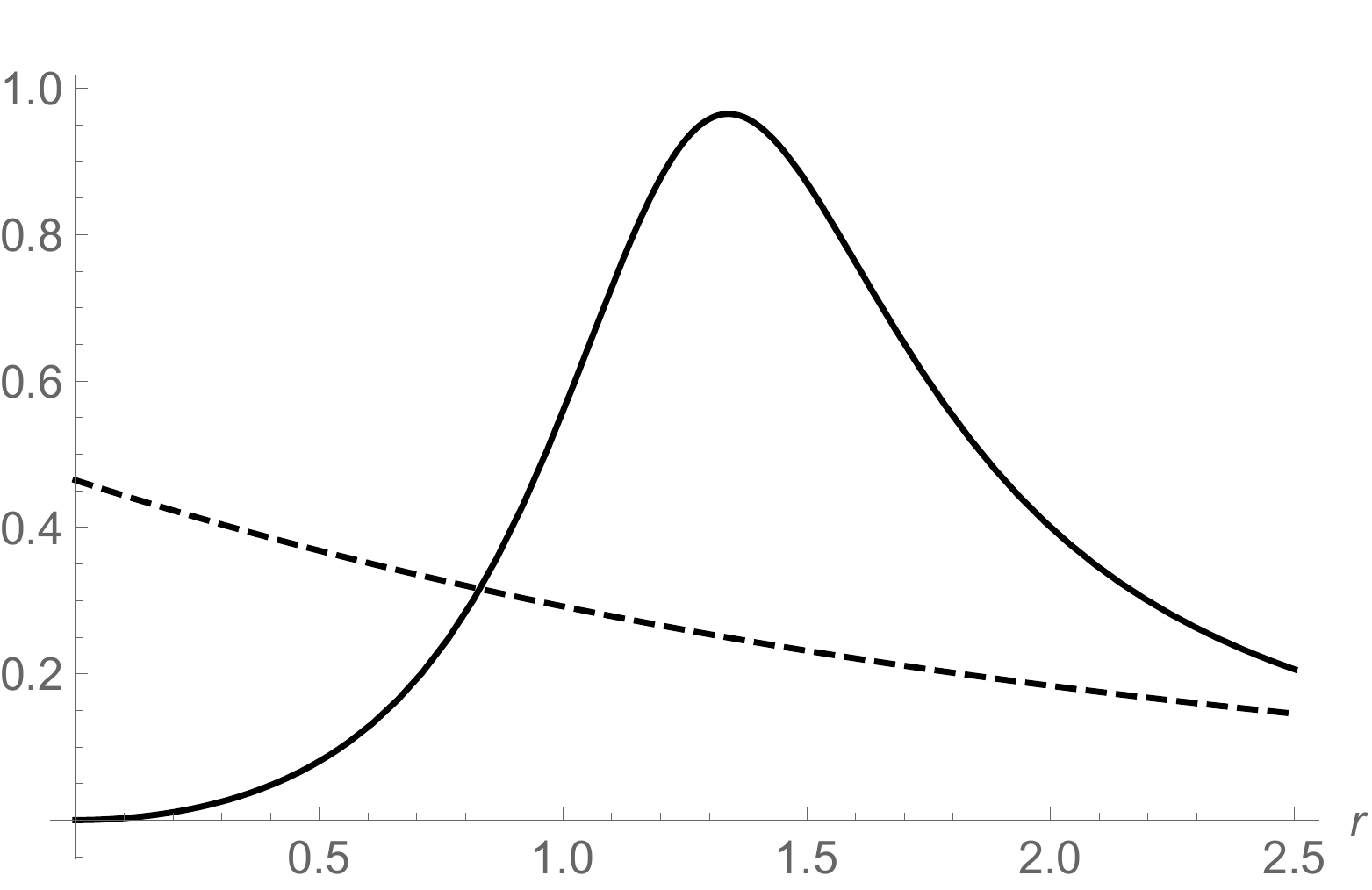}
$\quad$
\includegraphics[width=7.5cm,height=5cm]{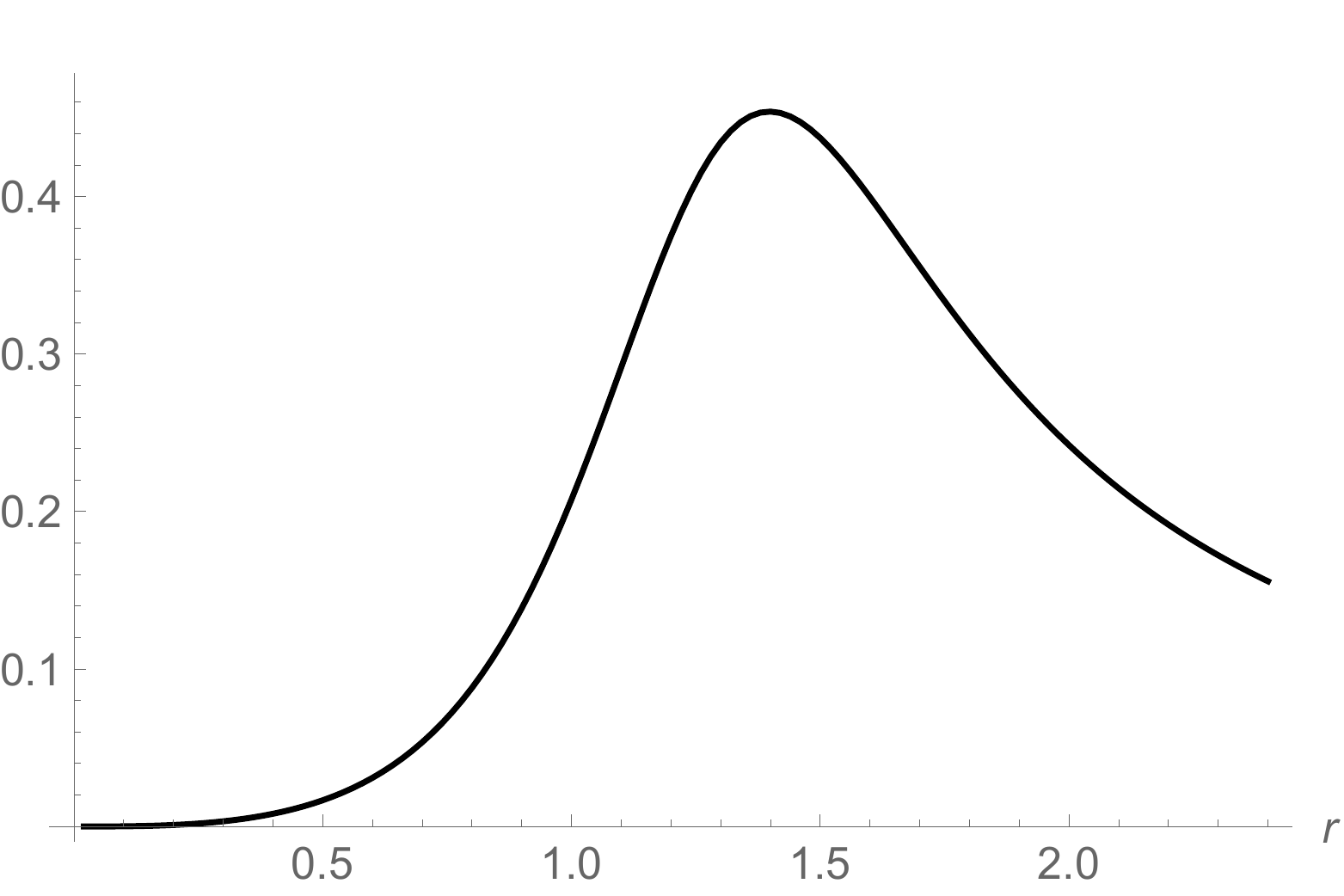}
\caption{Left panel: probability density $\mathcal{P}_{\rm H}$ in Eq.~\eqref{PhR}
for the horizon size (solid line) compared with the probability density $\mathcal{P}_{\rm S}$
of the resonance size (dashed line).
Right panel: probability density $\mathcal{P}_{<}$ in Eq.~\eqref{PrlessH} the resonance
is located within its gravitational radius of size $r$.
All lengths are in units of $\lp$.
\label{probDens}}
\end{figure}

First of all, we obtain the resonance wave-function in position space by
projecting $\psi_{\rm S}$ in Eq.~\eqref{BW} on the spherical Bessel function
\be
j_0(E,r)
=
\frac{\sin(E\,r)}{E\,r}
\ ,
\ee
that is
\be
\psi_{\rm S}(r)
&=&
\int_0^\infty
\psi_{\rm S}(E)\,j_0(E,r)\,\d E
\nonumber
\\
&\simeq&
\int_0^\infty
\frac{E^2\,\d E}
{E^2-M_{\rm BH}^2 +i\,M_{\rm BH}\,\Gamma_{\rm BH}}
\frac{\sin(E\,r)}{E\,r}
\nonumber
\\
&\simeq&
\frac{1}{r}\,
\exp\left[-i\,\frac{M_{\rm BH}}{\mpl}\,\sqrt{1-i\,\frac{\Gamma_{\rm BH}}{M_{\rm BH}}}\,\frac{r}{\lp}\right]
\ ,
\ee
where we omitted a normalization factor for simplicity.
We can then compute the probability density in Eq.~\eqref{PsR} for the resonance size,
the probability~\eqref{PrlessH} that the resonance is inside its own gravitational radius,
and the probability~\eqref{Pbh} that it is a black hole
\be
P_{\rm BH}(M_{\rm BH},\Gamma_{\rm BH})
\simeq 0.48
\ .
\label{Pbhp}
\ee
See also Fig.~\ref{probDens} for plots of the above quantities.
This result is interesting as it is compatible with the interpretation of the poles in the
resummed graviton propagators as black holes precursors.
If the probability had been much smaller than one, the interpretation as black hole
would have been challenged.
If it had been close to one, we would expect the black hole to be semi-classical but
this would be inconsistent with our expectation and model for the mass spectrum
described above.
A probability around one half is precisely what one would expect from a black hole
precursor.  
\subsection{Decay time}
The decay time $\tau$ of a common resonance can be estimated from the uncertainty relation
\be
\tau\,\Delta E
\simeq 
\lp\,\mpl
\ ,
\ee
and would be extremely short for our lightest black hole, namely
\be
\tau_{\rm BH}
\simeq 
\frac{\lp\,\mpl}{\Gamma_{\rm BH}}
\simeq\lp=\tau_{\rm p}
\ .
\label{life}
\ee
However, if the probability $P_{\rm BH}$ is significantly close to one, the resonance 
should decay more slowly.
Given the non-local nature of $P_{\rm BH}$, a precise estimate would require a numerical
analysis~\cite{timeEvo}, but we can obtain a rough estimate by simply considering that the
(initial) decay probability is reduced by $(1-P_{\rm BH})$, so that
\be
\tau_{\rm BH}
\simeq
\frac{\lp\,\mpl}{\Gamma_{\rm BH}\,(1-P_{\rm BH})}
\simeq 2\,\tau_{\rm p}
\ .
\ee
Again this result confirms the interpretation of the poles as black hole precursors,
since their lifetime is close to the Planck time.
\par
We emphasize that all estimates in this section were obtained by
assuming that  the proper mass and length scales in the definition of the gravitational
radius~\eqref{hoop} have their traditional values (i.e.~$\mpl \sim 10^{19}$~GeV).
This is consistent, since the coefficient of the Ricci scalar in the effective action is not
affected by the quantum corrections we have considered.
Here, we have not considered the running of the Planck mass, since there are not
many particles in the standard model, this would be a small effects~\cite{Calmet:2008tn,fixepoint,Anber:2011ut}. 
There are however well known models which can affect significantly the value of
the coefficient of the Ricci scalar.
For examples, models with a large extra-dimensional
volume~\cite{ArkaniHamed:1998rs,Randall:1999ee}.
Note that these models would not only affect the value of the coefficient of the Ricci scalar,
but also affect the effective theory itself and thus the resummed propagator calculation as well.
This effect would have to be carefully studied in these models.
\section{Conclusions}
\setcounter{equation}{0}
In this paper, we have studied the properties of the poles of the resummed graviton propagator obtained
by resumming bubble matter diagrams which correct the classical graviton propagator.
These poles had been interpreted as black holes precursors previously.
Here, we shown using the Horizon Wave-Function formalism that these poles indeed have properties
which make  them compatible with being black hole precursors.
In particular, when modeled with a Breit-Wigner distribution, they have a well defined gravitational radius.
The probability that the resonance is inside its own gravitational radius, and thus that it is a black hole
is roughly $50\%$.
The mass, width and gravitational radius as well as the existence of an horizon depends on the matter
content of the theory.
Here we have assumed that the particle content is that of the standard model of particle physics.
Our results confirm the previously proposed interpretation of these poles as black hole precursors.
\section*{Acknowledgments}
R.C. is partly supported by the INFN grant FLAG.
The work of X.C. is supported in part  by the Science and Technology Facilities Council (grant number  ST/J000477/1). 
\appendix
\section{Useful integrals}
\label{AppI}
\setcounter{equation}{0}
In order to compute integral of functions such as the one in Eq.~\eqref{PhR},
it is useful to define the dimensionless variables
\be
x+1
&=&
\frac{\mpl^2\,\rh^2}{4\,\lp^2\,M_{\rm BH}^2}
\nonumber
\\
\\
\gamma_{\rm BH}
&=&
\frac{\Gamma_{\rm BH}}{M_{\rm BH}}
\simeq
0.83
\ ,
\nonumber
\ee
and 
\be
\Lambda
=\frac{\mpl^2\,R_c^2}{4\,\lp^2\,M_{\rm BH}^2}-1
=
\left(\frac{E_c}{M_{\rm BH}}\right)^2-1
\simeq 3
\ .
\ee
We can then write
\be
F_n(\gamma_{\rm BH},\Lambda)
=
\int_{-1}^\Lambda
\frac{(x+1)^{\frac{n+1}{2}}}{x^2+\gamma_{\rm BH}^2}\,\d x
\ ,
\ee
and obtain, in particular,
\be
F_0(6/7,3)
&\simeq&
2.8
\\
F_1(6/7,3)
&\simeq&
3.5
\\
F_2(6/7,3)
&\simeq&
4.6
\ .
\ee
%
%
%
%
%

%
\end{document}